\documentclass[useAMS,usenatbib]{mn2e}
\usepackage{epsfig}

\newcommand{\Lsol}{L$_{\odot}$}
\newcommand{\Msol}{M$_{\odot}$}
\newcommand{\HI}{H\,{\sc {i}}~}

\newcommand{\Msold}{M$_{\odot}$\,yr$^{-1}$}
\newcommand{\Vlsr}{V$_{\rm lsr}$}
\newcommand{\Vexp}{V$_{\rm exp}$}

\newcommand{\Teff}{T$_{\rm eff}$}
\newcommand{\kms}{km\,s$^{-1}$}

\title[The \HI emission at 21 cm from X Her]{Atomic hydrogen in AGB circumstellar environments. 
A case study\,: X Her}
\author[E. Gardan, E. G\'erard and T. Le Bertre]{E. Gardan$^{1,3}$, E. G\'erard$^{2}$ and 
T. Le~Bertre$^{1}$\\
$^{1}$LERMA, UMR 8112, Observatoire de Paris, 61 av. de l'Observatoire,
           F-75014 Paris, France\\
$^{2}$GEPI, UMR\,8111, Observatoire de Paris, 5 place J. Janssen, 
           F-92195 Meudon Cedex, France\\
$^{3}$Observatoire de Bordeaux, 2 rue de l'Observatoire, F-33270 Floirac, France}
\begin{document}

\date{4 October 2005 (accepted)}

\pagerange{\pageref{firstpage}--\pageref{lastpage}} \pubyear{2005}

\maketitle

\label{firstpage}

\begin{abstract}
We report the detection of the \HI line at 21 cm from the circumstellar shell around the AGB star 
X Her using the position-switching technique with the Nan\c cay Radio Telescope. 
At the star position the line shows 2 components: (i) a broad one (FWHM $\sim$ 
13 \kms) centered at $-$72.2 \kms, and (ii) a narrow 
one (FWHM $\sim$ 4 \kms) centered at $\sim$ $-$70.6 \kms. 
Our map shows that the source associated to the broad component 
is asymmetric with material flowing preferentially towards the North-East. 
This source extends to $\sim$ 10\arcmin ~($\sim$ 0.4\,pc) from the star in that direction.
On the other hand, the narrow component is detected only at the star position 
and indicates material flowing away from the observer. 
The total mass of atomic hydrogen is $\sim$ 6.5 10$^{-3}$ \Msol ~which, within a factor 
2, agrees with the estimate obtained from IRAS data at 60 $\mu$m.

\end{abstract}


\section{Introduction}

Low and intermediate mass stars (1$<$M/\Msol$<$6-8) lose most of their mass 
during their evolution on the first red giant branch (RGB) and 
on the asymptotic giant branch (AGB). This phenomenon is known mainly 
from indirect arguments. One of the reasons is that mass loss develops irregularly 
on very different timescales, some of which may be short compared to 
the stellar evolution, e.g. down to a few years. It has therefore been 
difficult to establish a balance of the mass loss for the various types 
of stars. Furthermore most of the ejected matter is in the form of hydrogen. 
Although very abundant this element is difficult to detect in circumstellar 
shells. In any direction on the sky, the \HI line at 21 cm is dominated 
by galactic interstellar emission (Hartmann \& Burton 1997). 
On the other hand, the low excitation rotational 
lines of molecular hydrogen are in the infrared range (28 and 17 $\mu$m)
and difficult to observe from the ground.

Glassgold \& Huggins (1983, GH1983) have discussed the nature of circumstellar 
hydrogen. For stellar effective temperatures (\Teff) larger than 2\,500 K, hydrogen 
should be mainly in atomic form. On the other hand for \Teff $\le$ 2\,500 K  
hydrogen should be molecular in the upper atmosphere and in the inner circumstellar 
shell. Molecular hydrogen will eventually be photo-dissociated by the interstellar 
radiation field (ISRF) in the outer circumstellar shell, at distances of the order 
of typically 10$^{17}$ cm. 

The past attempts to detect the \HI ~line at 21 cm from mass losing red giants 
have failed, except on Mira (Bowers \& Knapp 1988). Nevertheless, after 
the renovation of the Nan\c cay Radio Telescope (NRT), we succeeded in detecting \HI 
from various circumstellar shells using a new observing technique  
(Le~Bertre \& G\'erard 2001; G\'erard \& Le~Bertre 2003, Paper I; Le~Bertre \& G\'erard 2004, 
Paper II). The reason is that this telescope is well adapted to the detection of extended 
low-level surface brightness sources and that we now systematically explore 
the spatial distribution of the emission by using the position-switching  
technique with different beam offsets. Indeed the circumstellar emission 
is extended and, for closeby sources, may reach a size of $\sim$ 1 degree 
over the sky (Paper II).

The emission line at 21 cm is a particularly useful tracer of circumstellar shells 
because its flux translates directly into a quantity of hydrogen (knowing the 
distance, which is relatively easy for AGB sources). 
Also hydrogen is a major component of red giant circumstellar shells 
so that the conversion to total mass is less liable to abundance ratio uncertainties. 
Furthermore, atomic hydrogen is not easily photo-ionized by the ISRF, and 
can be used as a probe of the most external parts of the shells where stellar winds 
interact with the Interstellar Medium (ISM). Indeed our \HI spectra on one source 
(Y CVn) show the effect of the slowing-down of circumstellar material 
by the surrounding ISM (Paper II). Young et al. (1993b) also found that the IRAS data 
at 60~$\mu$m imply a slowing-down in the outer shells of AGB stars.

In this paper, we show spatially resolved data on the environment of the red giant X Her. 
We have developed a model of \HI ~emission and compare its predictions with our data.

\section[]{Source Properties}

The position of the star X Her (IRAS 16011+4722) has been determined by 
Hipparcos (Perryman et al. 1997): $\alpha_{\rm 2000.0}$ = 16$^h$02$^m$39.17$^s$, 
$\delta_{\rm 2000.0}$ = +47$^{\circ}$14\arcmin25.28\arcsec, which translate to galactic 
coordinates: l$^{\rm II}$ = 74.46$^{\circ}$, b$^{\rm II}$ = 47.79$^{\circ}$. 
The proper motion, $-$68 and +64 mas yr$^{-1}$ in equatorial coordinates, is towards 
the North-West. 

\subsection{stellar properties}

The star is a long-period semi-regular variable (SRb) with a period 
of 95.0 days (General Catalogue of Variable Stars). 
Light variations are small ($\Delta$V $ \sim \pm$ 0.5 mag.) and irregular.
A period analysis of V and I$_{\rm C}$ photometric data over 1600 days (Lebzelter 
\& Kiss 2001) yields a main period of 101 days and several ill-defined longer ones.
It has a variable spectral type, from M6 to M8. Dumm \& Schild (1998)
estimate the stellar effective temperature at 3\,161\,K, and Dyck et al. 
(1998) at 3\,281\,$\pm$\,130\,K. Therefore we adopt \Teff ~$\approx$ 3\,200\,K and, 
according to GH1983, atomic hydrogen should be the dominant species in the stellar atmosphere. 
The parallax has been measured by Hipparcos, 7.26\,$\pm$\,0.70\,mas. 
In the following we adopt a distance of 140~pc. 

For a K magnitude of $-$1.42 (Jura \& Kleinmann 1992) and a red-giant 
bolometric correction of 2.7 (Le Bertre et al. 2001), the luminosity 
should be around 4\,800 \Lsol. This clearly places X Her on the AGB. 
However it should not be strongly evolved as searches for technetium 
failed (Little et al. 1987; Lebzelter \& Hron 1999).
 
Velocity variations have been detected by Hinkle et al. (2002) in the CO 
lines at 1.6 $\mu$m with a possible period of 660 days and an average 
heliocentric velocity of $-$90.3$\pm$0.2 \kms, which translates to \Vlsr\,=\,$-$73.1 \kms.

\subsection{circumstellar shell properties} 

\begin{table*}
 \centering
  \caption{Observational results of X Her molecular radio line emissions taken from the literature.}
  \begin{tabular}{llllll}
  \hline
line    &  \Vlsr  & \Vexp  &  \Vlsr  & \Vexp  & Reference\\
 \hline
CO 2-1  &  $-$72.8 $\pm$ 0.8  &  8.5 $\pm$ 1.0  & $-$73.2 $\pm$ 0.4 &  3.2 $\pm$ 0.5 & 
Knapp et al. (1998)\\
CO 3-2  &  $-$73.2 $\pm$ 0.5  &  9.0 $\pm$ 1.0  & $-$73.1 $\pm$ 0.3 &  3.5 $\pm$ 1.4 & 
Knapp et al. (1998)\\
CO 3-2  &  $-$73.0      &  8.0      & $-$73.1     &   3.1  & Kerschbaum \& Olofsson (1999)\\
SiO 2-1 &  $-$73.0      &  8.0      & $-$72.0     &   2.2  & Gonz\'alez-Delgado et al. (2003)\\
\hline
\end{tabular}
\label{CO_SiOtab}
\end{table*}

X Her has an infrared excess indicating that it is undergoing mass loss. 
The IRAS LRS spectrum (LRS class 24) and the ISO SWS spectrum show the amorphous 
silicate features at 10 and 18 $\mu$m and an unidentified feature at 13 $\mu$m possibly due to 
some crystalline form of alumina (Sloan et al. 2003). The IRAS data at 60 $\mu$m
show that the infrared source is extended with a diameter of 12.4$'$ or 0.5 pc 
(Young et al. 1993a).

CO emission in the direction of X Her was discovered by Zuckerman \& Dyck (1986).
High-quality CO spectra reveal that the (2-1) and (1-0) lines have a 
complex profile with drastic variations of line shape with position 
(Kahane \& Jura 1996, KJ1996). There is a narrow component 
centered at $-$73.3 \kms ~of width about 5 \kms, plus a blue shifted wing 
extending out to $-$82 \kms ~and a red-shifted wing extending out to $-$64 \kms. 
The red wing is noticeably more intense than the blue wing. 
The narrow feature is centered on the star position whereas 
the strong red wing is found to the North-East of the 
star, and the weak blue wing to the South-West. KJ1996 interpret these CO profiles 
as resulting from three elements in the circumstellar shell: a very slowly expanding 
spherical wind (\Vexp\,= 2.5 \kms) and two higher velocity ones 
(\Vexp ~$\geq$ 10 \kms). The latters are assumed to be parts of 
a weakly collimated bipolar outflow whose axis is inclined at 
$\sim$ 15$^{\circ}$ to the line of sight.

Knapp et al. (1998, K1998) obtained high-quality high-resolution ($\leq$ 0.2 \kms) CO (2-1) 
and CO (3-2) line-profiles of X Her. The CO (2-1) profile is very similar to that obtained 
by Kahane \& Jura (1996), whereas the CO (3-2) one is also composite, but symmetric. 
Knapp et al. find that X Her belongs to a class of stars 
that show composite profiles with a narrow feature superimposed on a broad one. 
They interpret this kind of profile as evidence for the presence of 
two steady winds from the central star, the narrow component indicating the 
onset of a new phase of mass loss. 
Assuming spherical symmetry, they estimate the velocity of the slow wind at 3.4 \kms 
~and of the fast wind at 9.0 \kms, and the mass loss rates at 3.4 10$^{-8}$ and 
1.1 10$^{-7}$ \Msold, respectively. Olofsson et al. (2002) obtain similar results, 
but the expansion velocities are smaller (2.2 and 6.5 \kms, respectively) 
because they account for the effect of turbulent broadening.

Nakashima (2005) has produced a map of the circumstellar environment of X Her 
in the CO (1-0) line with the BIMA interferometer. Like KJ1996 he finds that 
the red-shifted emission is offset to the North-East, and the blue-shifted emission,
to the South-West. He estimates the position angle of the bipolar structure axis 
at 61$^{\circ}$ in the plane of the sky. The structure associated to the narrow 
spectral feature is tentatively ascribed to a rotating disk.

Gonzalez-Delgado et al. (2003) have observed the SiO (2-1) thermal emission from X Her. 
Like for CO, the line-profile is composite with 2 components whose central velocities and widths 
agree with the CO ones (Table \ref{CO_SiOtab}). 

X Her was detected neither in the OH maser main lines (Lewis et al. 1995), nor as an H$_2$O maser 
at 22 GHz (Lewis 1997).

\section[]{Observations}

The NRT is a clear aperture radio-telescope with a tiltable flat  
reflector illuminating a fixed sphere. The aperture is rectangular with effective dimensions 
160\,m$\times$30\,m (as long as the declination is smaller than 53$^\circ$, which is the 
case of X Her). The beam has thus 
a HPBW of 4$'$ in right ascension and 22$'$ in declination. 
The point source efficiency is 1.4 K\,Jy$^{-1}$ at 21 cm  
and the beam efficiency, measured on the Moon, 0.65. 
Sources are tracked for about one hour around 
meridian by moving a focal carriage bearing the receivers 
(the main collecting fixed mirror being over-sized in right ascension). 
Stray radiation and side-lobes have been minimized 
through a careful design provided by CSIRO (van Driel et al. 1996; 
Granet et al. 1999) and can 
be readily evaluated by examining elementary scans obtained at different 
hour angles. Drift scans obtained on radio-continuum point sources show 
that the beamprofile is as expected from an unobstructed rectangular 
aperture, with secondary lobes $<$ 5\%. 
However, internal reflections between 
the horn and the spherical mirror may occur and produce an oscillation 
in the spectra with a period of 536 kHz (or 113 \kms ~at 1420 MHz). 
This artefact, usually called ``specular reflection'', 
is exactly removed when using the position-switch 
mode of observation, whereas, in the frequency-switch mode, it can be kept 
down to 0.15 K by selecting an offset equal to a multiple of 536 kHz.
In principle, there is no stray radiation for sources with declination 
smaller than 53$^\circ$, above which the tiltable plane mirror starts 
to diaphragm the main beam of the fixed spherical mirror. However, we may 
get direct spill-over from the sky around the spherical primary reflector 
due to an incomplete apodisation inside the focal system. 
Continuum emission is perfectly removed in both modes, 
but line emission, if any, may remain when using the frequency-switch mode. 

\begin{figure}
\centering 
\epsfig{figure=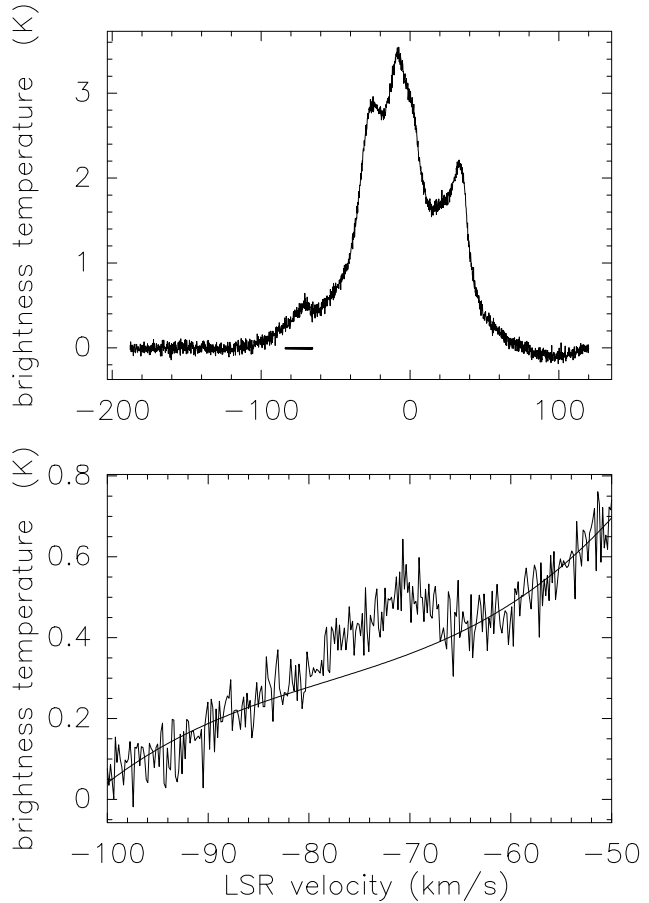,angle=0,width=8.8cm}
\caption[]{
{\it Upper panel:} 
\HI on-source spectrum obtained in f-switch mode; the horizontal bar marks the velocity 
spread corresponding to the X~Her CO emission (from $-$83 to $-$65 \kms; K1998).
A sinusoid of period 113 \kms ~and amplitude 0.10 K has been 
subtracted from the raw spectrum (see text).
{\it Lower panel:} 
Detail from --100 to --50 \kms ~with a third-order baseline 
to isolate the X Her emission.
}
\label{fswitch}
\end{figure}

For the observations we adopted the same approach as for EP Aqr and Y CVn (Paper II). 
A frequency-switch (f-switch) spectrum was first acquired 
on the source (as defined by the star optical position) 
to estimate the \HI background emission and thus check 
the feasibility of the project. Galactic \HI 
emission is clearly detected at \Vlsr $> -$50 \kms ~(Fig.~\ref{fswitch}).
The galactic emission stays below 0.5~K in the velocity range expected for X Her.  
The \HI emission from X Her itself can be suspected directly on this f-switch spectrum 
between $-$80 and $-$60 \kms ~as an excess of about 0.1 K around $-$70 \kms.

However the Leiden-Dwingeloo ``Atlas'' of Hartman \& Burton (1997) shows \HI emission at 
$\sim$ 0.6$^{\circ}$ North-West of X Her in the range $-$100 to $-$60 \kms. 
The Dwingeloo telescope has a HPBW of 36$'$ and the Atlas data 
were observed at 0.5$^{\circ}$ 
spacings in both galactic coordinates.
This cloud (hereafter Cloud I, 
l$^{\rm II}$ = 75.0$^{\circ}$, b$^{\rm II}$ = 47.5$^{\circ}$) is 
responsible for a contamination of our data North of X Her. On this Atlas
one sees also a second source (Cloud II, 
l$^{\rm II}$ = 77.0$^{\circ}$, b$^{\rm II}$ = 48.5$^{\circ}$), 
but further to the North-West ($\sim 2^{\circ}$). 
These 2 sources seem aligned with a large arc-shaped structure that is above the galactic 
plane and stretches from l$^{\rm II} \sim 200^\circ$ to l$^{\rm II} \sim 80-100^\circ$. 
However both clouds are compact ($\phi \sim 1^{\circ}$) and seem more 
likely related to the compact high-velocity clouds (CHVCs) discussed by de Heij et al. (2002). 
CHVCs are compact and isolated on the sky with a core-halo structure.
The X Her position does not fall exactly on one of the Atlas grid points, but \HI emission,  
at the same velocity as seen in Fig.~\ref{fswitch}, is detected at a 0.1 K level on the 
nearest point. This emission may come from a diffuse halo associated to Cloud I as those 
detected by de Heij et al. around several CHVCs. 

Incidentally a feature at +40 \kms, that the Leiden-Dwingeloo Atlas do not show, 
is visible in our spectrum. It is an artefact due to galactic \HI emission, around
$\sim$ 16$^{h}$ in right ascension and $\sim$ --35$^{\circ}$ in declination, coming directly 
into the focal system from around the spherical primary mirror. We have checked that 
this artefact is perfectly removed when using the position-switching technique. 

\begin{figure} 
\centering
\epsfig{figure=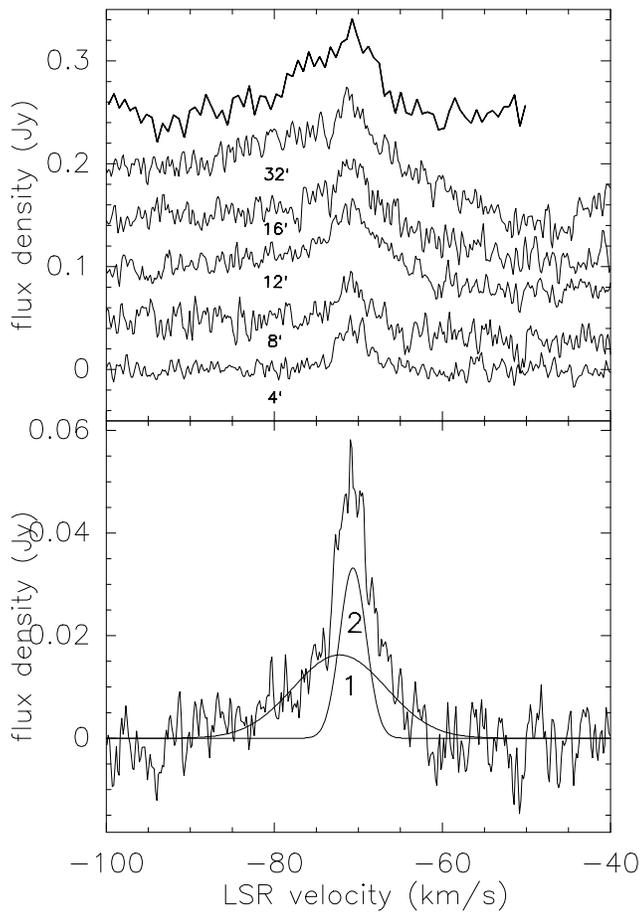,width=8.8cm}
\caption[]{
{\it Upper panel:} 
Spectra obtained in the position-switch mode with the source centered 
(``on'') and the off-positions taken in the East-West direction 
at $\pm$4\arcmin, 
$\pm$8\arcmin, $\pm$12\arcmin, $\pm$16\arcmin ~and $\pm$32\arcmin 
~(thin lines), and baseline-subtracted f-switch spectrum (thick line). 
For clarity the spectra are successively shifted upwards by 0.05 Jy 
and labelled with the corresponding offsets. 
{\it Lower panel:} average of the 3 position-switch spectra ($\pm$4\arcmin, 
$\pm$8\arcmin, $\pm$12\arcmin) presented in the upper panel. 
}
\label{Spect_Centre_pos_switch}
\end{figure}

The spectra obtained in the position-switch mode with the star placed in the central 
beam are presented in Fig.~\ref{Spect_Centre_pos_switch}. 
The spectral resolution corresponds to 0.16\,\kms. 
The off-positions are taken at $\pm$ n NRT beams in the East-West 
direction, with n = 1, 2, 3, 4 and 8. The symmetrical off-positions are 
averaged and subtracted from the spectrum obtained on the source, yielding 
a source spectrum corrected from an underlying background interpolated 
successively between $\pm$4$'$, $\pm$8$'$, $\pm$12$'$, 
$\pm$16$'$ and $\pm$32$'$. This procedure is efficient in removing 
the galactic \HI background if it varies linearly between the 2 off-positions. 
However it also removes genuine emission when the source is extended: 
position-switch observations with larger throws are then needed, at the 
expense of a larger interstellar confusion arising from the quadratic variation 
of the background \HI intensity.

The position-switch profiles can be decomposed 
in 2 components: (i) a broad (FWHM $\sim$ 13 \kms) feature centered at $-$72.2 \kms ~(Comp. 1), 
and (ii) a narrow (FWHM $\sim$ 4 \kms) one centered at $-$70.6 \kms ~(Comp. 2).
The gaussian fit to Comp. 1 may have been slightly biased towards a higher velocity than 
real, due to the asymmetric brightness distribution (see below) and to the noise in the data. 
Its velocity can therefore be considered as consistent with the other radio lines 
(Table~\ref{CO_SiOtab}), 
and with the star radial velocity estimated by Hinkle et al. (2002). On the other hand 
Comp. 2 is clearly red-shifted by about 2-3 \kms. The narrow component is present on all
position-switch spectra with the same intensity. Therefore it corresponds to a compact 
unresolved source ($\phi < 4$\arcmin). The broad one (Comp. 1) is hardly present at $\pm$ 1 beam 
and grows as the beam throw increases. Its intensity reaches a maximum at $\pm$ 3 beams
and stays constant beyond. 
This is confirmed by the comparison with the f-switch spectrum displayed 
at the top of Fig.~\ref{Spect_Centre_pos_switch} for which a third-order baseline
fitted to the $-$100, $-$83 \kms ~and $-$65, $-$50 \kms 
~ranges has been subtracted (see Fig.~\ref{fswitch}, lower panel). 
The broad component is detected on this baseline-subtracted 
f-switch spectrum with the same centroid velocity and the same intensity  
as in the $\pm$12$'$ position-switch spectrum. 
While our symmetric East-West position-switching technique is 
efficient in removing the linear gradient of the background \HI emission, the quadratic term 
stays and is no longer negligible at large beam throws. This is particularly clear on 
Fig.~\ref{Spect_Centre_pos_switch} (top) at $\pm$32\arcmin ~where extraneous wings appear 
in emission on the blue side and in absorption on the red side of the X Her profile. 
The negative signature around $-$60 \kms ~is probably connected 
to the rising main galactic \HI emission (see  Fig.~\ref{fswitch}). 
For the display in the lower panel of Fig.~\ref{Spect_Centre_pos_switch} we have averaged 
only the spectra up to $\pm$ 3 beams. From these position-switch spectra 
one can estimate a source size $\phi \leq 20$\arcmin.

\begin{figure}
\centering
\epsfig{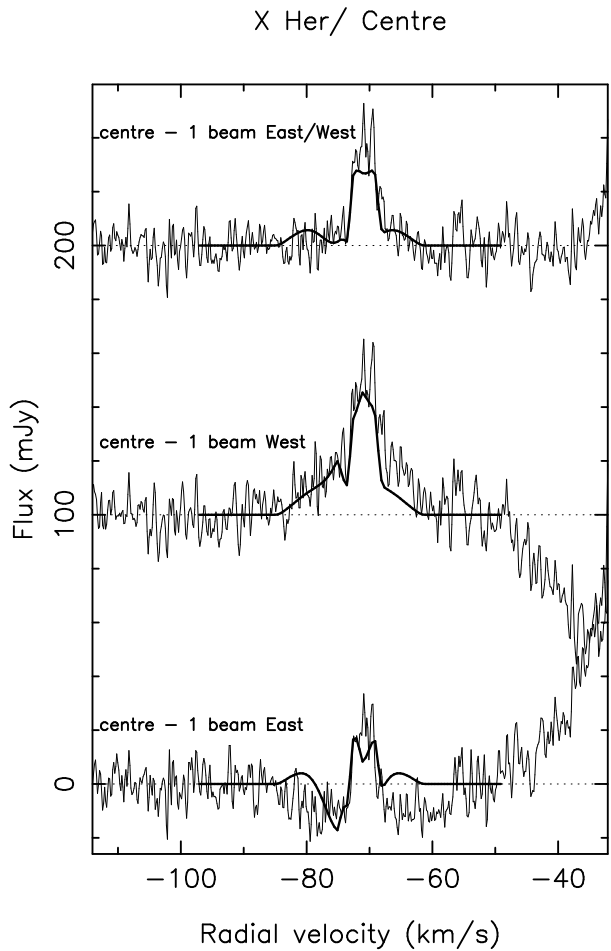}
\caption{X Her position-switch spectra (thin lines) and modelled spectra 
(Sect. 5.2; thick lines). From bottom to top\,: (i) the off-position is taken at +4\arcmin, 
(ii) the off-position is taken at $-$4\arcmin, (iii) the off-positions at + and $-$4\arcmin 
~are averaged.}
\label{fluxC1E_C1W_C1EW_model_1_2}
\end{figure}

However, the brightness distribution is very asymmetric as can be seen on 
Fig.~\ref{fluxC1E_C1W_C1EW_model_1_2} where the reference spectra at $\pm$ 1 beam East 
and West are subtracted separately. One notes that Comp. 1 is extended East and that 
it is even more intense at 1 beam East than on the star position. 

\begin{figure}
\centering 
\epsfig{figure=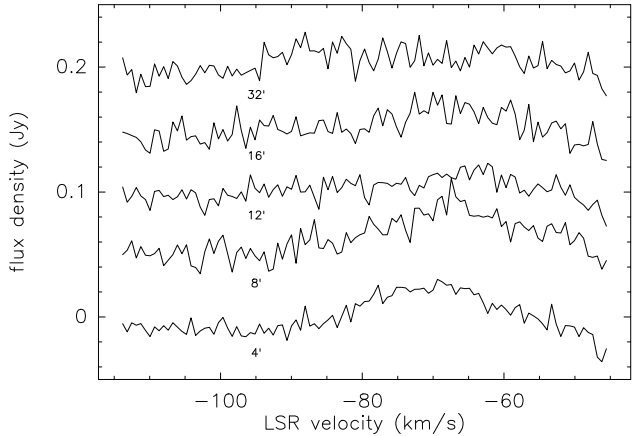,angle=270,width=8.8cm}
\caption[]{Differences between the two reference spectra 
obtained at $\pm$4\arcmin, $\pm$8\arcmin, $\pm$12\arcmin, 
$\pm$16\arcmin ~and $\pm$32\arcmin ~in RA from X Her. 
For clarity the spectra are successively shifted upwards by steps 
of 0.05 Jy and labelled with the corresponding offsets.}
\label{Spect_O_Centre}
\end{figure}

The differences between the 2 reference spectra (n``East''$-$n''West'') confirm 
these results: Comp. 1 is visible at 1 and 2 beams East and Comp. 2 is absent 
in all the reference spectra (Fig.~\ref{Spect_O_Centre}). No emission is detected 
West of the central position.
We stress that, in contrast to what we have done in Paper II, these differences 
are not normalized. One can note that the galactic \HI confusion at the position 
of X Her and in the velocity range $-$120 to $-$60\,\kms ~is very low. 
However its effect starts to be seen above $\sim$ $-$50 \kms ~at small beam throw ($\leq$12\arcmin) 
and extends to $-$95 \kms ~at $\pm$32\arcmin, although weak.

\begin{figure} 
\centering
\epsfig{figure=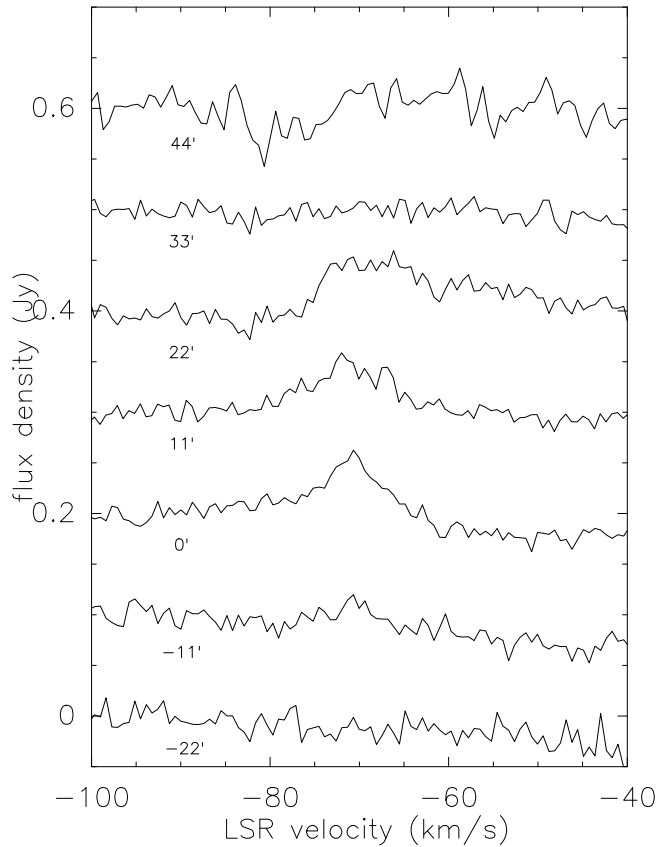,width=8.8cm}
\caption[]{From top to bottom: position-switch spectra with 
the central beam placed at $+$44\arcmin ~in declination, $+$33\arcmin, 
$+$22\arcmin, $+$11\arcmin, on source, $-$11\arcmin ~and  $-$22\arcmin.
The off-positions are taken at $\pm$12\arcmin ~(3 beams). For clarity 
the spectra are successively shifted by steps of 0.1 Jy and 
labelled with the corresponding positions in declination.
}
\label{Spect_C_3E_delta}
\end{figure}

We have also explored the \HI brightness distribution in the North-South direction 
(Fig.~\ref{Spect_C_3E_delta}) by steps of 11\arcmin ~(i.e. 1/2 beam). There is no emission 
South of X Her. Comp. 1 is detected at 1/2 beam North and probably also at 1 beam North.
This indicates that Comp. 1 is offset towards the North. Unfortunately, 
the northern spectra are contaminated by emission around $-$80 \kms.

\begin{figure*}
\centering
\epsfig{figure=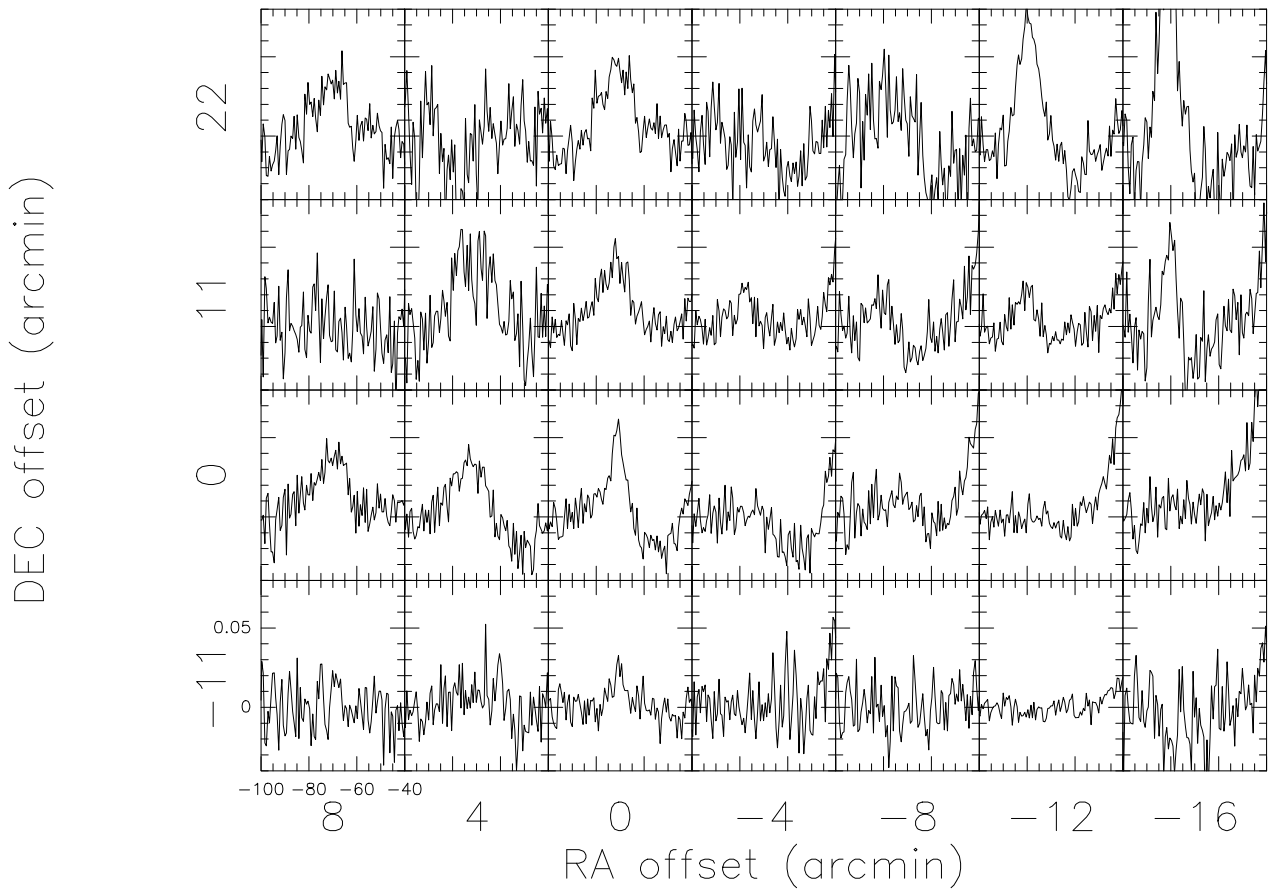,angle=270,width=19cm}
\caption[]{
Map of the 21 cm \HI ~emission from the X Her circumstellar envelope 
observed with the NRT. The steps are 4\arcmin ~in RA (1~beam) and 11\arcmin 
~in declination (1/2 beam). North is up and East to the left. The star position 
corresponds to the second row from the bottom, third column from the left.
Note the emission from Cloud I at $\sim -$80\,\kms ~in the upper right corner (North-West).}
\label{Map}
\end{figure*}

In Fig.~\ref{Map} we present a ``map'' of the \HI emission around X Her. 
To construct this map we have used the reference spectra obtained at 3 beams 
East of X Her, because the East-West extent of the source is limited to $\pm$10\arcmin   
~and because the reference spectra West of X Her are contaminated by Cloud I. 
Noteworthily, this cloud shows up clearly around $-$80\,\kms ~in the North-West corner of the map. 
Also the negative feature around $-$60 \kms ~visible on the star position 
and at 4$'$ East is an artefact due to the non-linear variation of the underlying galactic 
background. 
The emission associated to Comp. 1 is present at the star position and towards East and 
North. Comp. 2 is visible only on the spectrum obtained at the position of X Her and 
possibly on that done at 1/2 beam North. Finally we have computed the total \HI mass 
in the envelope of X Her by integrating the flux density under the profiles 
displayed in Fig.~\ref{Map}, restricted to $\pm$8\arcmin ~East-West 
and $\pm$11\arcmin ~North-South. We find a hydrogen mass of 6.5 10$^{-3}$ \Msol. 

\section[]{Description of the Envelope\\* Model}

In order to guide the interpretation of our spatially resolved spectra we have 
performed numerical simulations of the 21 cm (1420 MHz) emission
from an \HI circumstellar envelope. We assume that the matter is 
flowing radially from the central star. Because h$\nu$ $\ll$ kT,  
the brightness temperature is assumed to be directly proportional to the \HI column density. 
This hypothesis is valid as long as the hydrogen temperature is larger than 10 K. 
The emission is also supposed to remain optically thin ($\tau \ll$ 1). For a constant 
\HI mass loss of 10$^{-6}$ \Msold ~and a constant expansion velocity 
of 5 \kms, this hypothesis would break down at 10$^{16}$ cm from the central star, 
or 0.08$'$ at 140 pc (i.e. much less than the NRT beam size). 

The emission from the shell is convolved with the telescope response. First, in order 
to perform simple checks or tests, we adopt a constant response within an elliptical 
beam of minor axis 4$'$ in the East-West direction  and major axis 22$'$ in the North-South 
direction (hereafter ``boxcar'' response). Second, for a better fit to 
the observations, we adopt the response of a rectangular aperture 
which is given by the product:
\begin{equation}
{R(x,y) = \bigg( \frac{sin\,x}{x}\bigg)^{2}\times\bigg( \frac{sin\,y}{y}\bigg)^{2}}
\end{equation}
normalized such that the FWHM is 4$'$ in right ascension and 22$'$ in declination
(hereafter ``sinc'' response).

\subsection{spherical geometry}

\begin{figure}
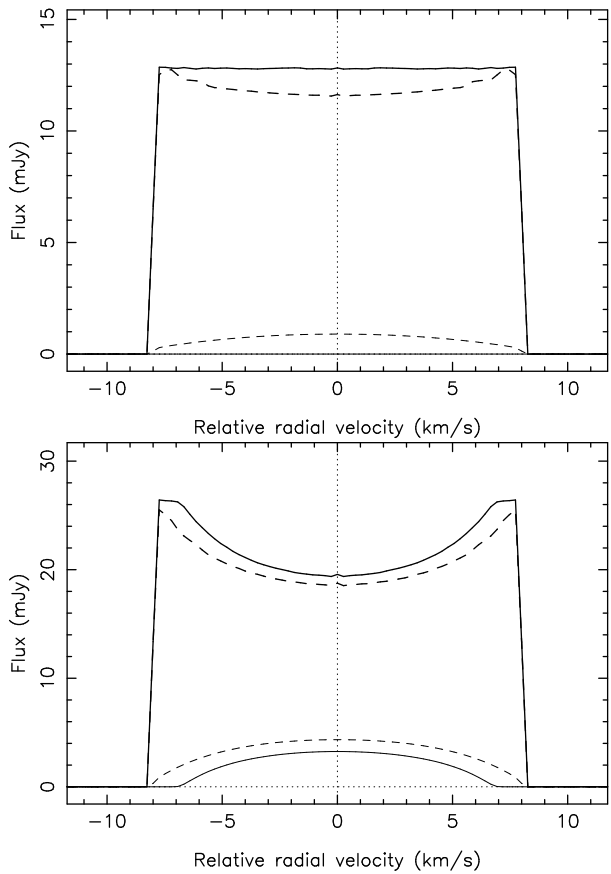

\centering
\epsfig{figure=Fig_Vconst_sphere_nonresolved_C_1.ps,angle=270,width=8.0cm}
\epsfig{figure=Fig_Vconst_sphere_resolved_C_1.ps,angle=270,width=8.0cm}
\caption{Simulation of the line-profiles produced by a source undergoing mass loss 
at a constant rate (1.0 $\times 10^{-7}$ \Msold ~in atomic hydrogen), 
at a constant expansion velocity (8.0 \kms) and located at 140 pc. 
The geometry is spherical. Continuous lines: 
boxcar response, dashed lines: sinc response (see text). 
The thick curves correspond to the source centered in the beam; the thin curves 
correspond to the source at an offset of 4$'$ in the East-West direction.
{\it Upper panel:} the inner radius is 0.1$'$ and the outer radius, 2.0$'$. 
{\it Lower panel:} same inner radius and an outer radius of 4.0$'$}.
\label{ModspherVconst}
\end{figure}

For a spherically symmetric shell, the density and the velocity depend only on r, 
the distance to the central star. In Fig.~{\ref{ModspherVconst}} (top) we show 
the results of our model for a source with a constant expansion velocity and unresolved 
by the telescope beam ($\phi$ = 4$'$). As expected, the line profile is rectangular 
for the boxcar response and no flux is detected at the position offset by 4$'$ in 
the East-West direction. When the more realistic sinc response is considered, the centre 
of the profile is depressed and the missing flux is obtained at an offset of 4$'$ from 
the sidelobe of the beam profile. If the source is extended ($\phi$ = 8$'$, bottom), 
a double-horn profile is obtained. The absence of such profiles in our data indicates that 
the velocity is not constant in the regions probed by our observations (cf. the discussion 
in Paper II).

\begin{figure}
\centering
\epsfig{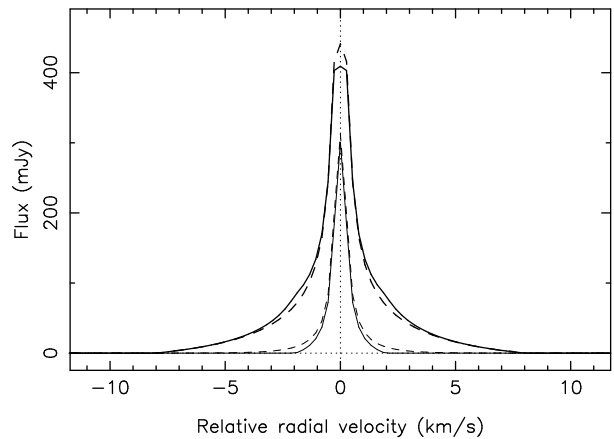}
\caption{Simulation of the line-profiles produced by a source undergoing mass loss at 
a constant rate (1.0 $\times 10^{-7}$ \Msold ~in H\,{\sc {i}}) and located at 140 pc. 
The geometry is spherical; the inner radius is 0.1$'$ and the outer radius, 4.0$'$. 
The velocity decreases linearly from the inner radius (8 \kms) to the outer one (0.2 \kms).
Continuous lines: boxcar response, dashed lines: sinc response (see text). 
The thick curves correspond to the source centered in the beam; the thin curves 
correspond to the source at an offset of 4$'$ in the East-West direction.}
\label{ModspherVdeclin}
\end{figure}

In Fig.~{\ref{ModspherVdeclin}} we show the line profiles obtained for a model with 
a velocity decreasing linearly with r, the mass loss rate in \HI being kept constant. 
The ISM will unavoidably slow down the expansion velocity once the densities 
become comparable (e.g. Young et al. 1993b).
The velocity and mass loss rate laws are arbitrary as the purpose is only 
to illustrate the effect of a velocity gradient on the line-profile. The time 
to build such an hypothetical shell would be 73.1 10$^3$ years, as compared 
to 9.5 10$^3$ years in the V=constant case (Fig.~{\ref{ModspherVconst}}, bottom). 

\subsection{non-spherical geometry}

As there is evidence from our \HI data and from CO rotational lines data 
that the X Her shell is not spherically symmetric, we have generalised the geometry 
of the model to the axi-symmetric case. 
The density and the velocity are defined in a source reference frame 
that can be orientated in any direction with respect to the line of sight (but 
we keep the hypothesis that the velocity is radial). We have 
checked on spherical cases that we find the same results as in the previous 
Section (although with considerably increased computing time). 

\begin{figure}
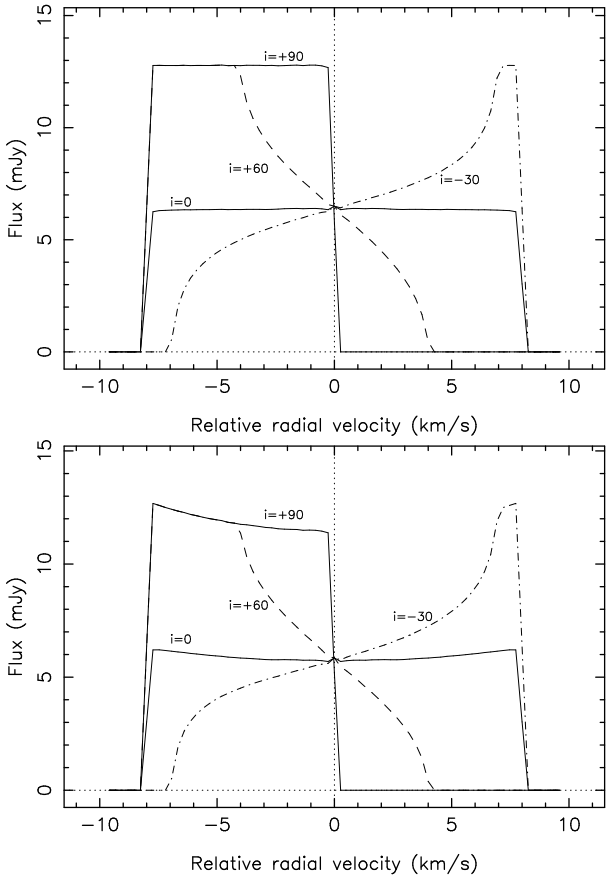

\centering
\epsfig{figure=Fig_Vconst_hemisphere_nonresolv_boxcar_cor.ps,angle=270,width=8.0cm}
\epsfig{figure=Fig_Vconst_hemisphere_nonresolv_sinc_cor.ps,angle=270,width=8.0cm}
\caption{Simulation of the line-profiles produced by a source undergoing mass loss 
within 2 $\pi$ steradian at a constant rate (0.5 $\times 10^{-7}$ \Msold), 
at a constant expansion velocity (8.0 \kms) and located at 140 pc. 
The source is a hemisphere whose axis is orientated at an angle i 
with respect to the plane of the sky (see text). 
The corresponding inner radius is 0.1$'$ and outer radius, 2.0$'$.
{\it Upper panel:} boxcar response. {\it Lower panel:} sinc response.}
\label{ModdemispherVconst}
\end{figure}

As an example we give in Fig.~{\ref{ModdemispherVconst}} the results from a 
hemispheric source with the same parameters as those used for the top panel 
of Fig.~{\ref{ModspherVconst}. The centre of the corresponding sphere is placed 
at the centre of the beam.
The axis of the hemisphere is inclined by an 
angle i with respect to the plane of the sky (the axis is in the plane of the sky 
and pointing to the West for i\,=\,0$^{\circ}$). 
The flux is one half of that obtained from a complete sphere.
One notes that in the boxcar 
case the profile is rectangular for i\,=\,0$^{\circ}$ and for i\,=\,$+/-$90$^{\circ}$; 
in the latter cases the centroid velocity is shifted by $-/+$\Vexp/2{\footnote{More generally 
a rectangular profile is obtained for an unresolved circular cone whose axis is perpendicular
to the plane of the sky.}}.

\section[]{Application to X Her}

In the following, we adopt a ``sinc'' response with parameters corresponding to the NRT 
at 21 cm. We also adopt a stellar radial velocity of $-$73.1 \kms ~(Hinkle et al. 2002).

\subsection{spherical model}

The source is assumed to have 2 shells with outflow velocities decreasing from the central 
star (Paper II). The two shells are invoked to explain the spectrally broad component 
which is spatially resolved (Comp. 1), and the spectrally narrow component 
which is not spatially resolved (Comp. 2).

For the inner shell we adopt an internal radius corresponding to 0.1\arcmin ~and an external 
one, to 1\arcmin. The velocity is decreasing linearly with radius from 3 \kms ~to 1 \kms. 
The value 3 \kms ~is a compromise between the estimates of K1998 and KJ1996. 
The flux of matter is kept constant and corresponds to 0.3~10$^{-7}$\,\Msold ~in
atomic hydrogen, in agreement with the estimate of K1998 for the total mass loss 
rate of the slow wind.

For the outer shell we adopt an internal radius of 2\arcmin ~and an external 
one of 10\arcmin. The external limit is consistent with our map (Sect.~3) and with the 
estimate of 6.2\arcmin ~obtained by Young et al. (1993a) from IRAS data at 60~$\mu$m.
The velocity is selected to decrease from 10 to 2 \kms. The starting value corresponds
to the estimate obtained from the CO profiles by K1998 and KJ1996. The flux of matter 
is set to 0.7~10$^{-7}$\,\Msold ~in \HI. 

\begin{figure}
\centering
\epsfig{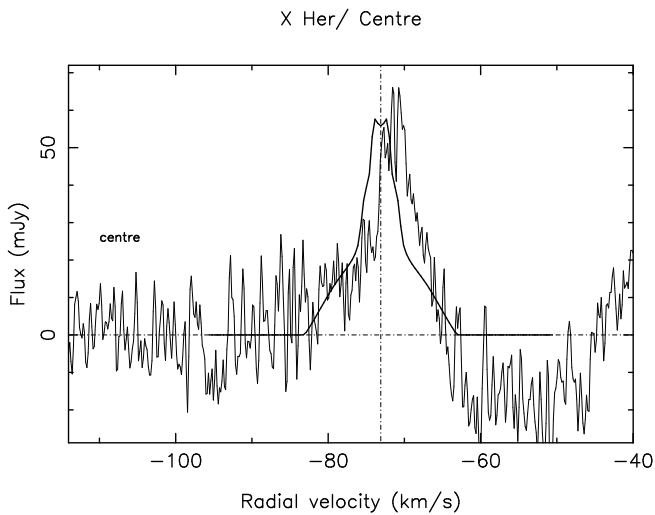}
\caption{\HI spectrum obtained with the NRT on the star position (thin line) and 
modelled spectrum with a spherical geometry (Sect. 5.1; thick line). The vertical 
line indicates the star radial velocity (\Vlsr = $-$73.1 \kms).}
\label{fluxCentre_C_model_spher}
\end{figure}

The fit to the \HI line profile obtained on the central position 
(Fig.~\ref{fluxCentre_C_model_spher}) is almost satisfactory, although Comp. 2 
is clearly red-shifted by about 2 \kms ~with respect to the model narrow emission 
coming from the inner shell. Furthermore, as expected, the profiles obtained 
away from the star position (not shown here) are not correctly reproduced. 
The model gives too much flux West and too little East. The same effect is noted 
in the North-South direction. 

\subsection{non-spherical model}
  
The emission associated to Comp. 1 can be understood with a source 
(Source (1)) which is roughly symmetric with respect to the plane of the sky 
because the centroid velocity corresponds to the star radial velocity.
But it has to be strongly weighted towards the North-East. 
We adopt a circular cone with an opening angle of 2$\times$75$^{\circ}$ 
and axis in the plane of the sky (i\,=\,0$^{\circ}$) 
at a position angle, PA\,=\,45$^{\circ}$. The inner limit is defined by 
a sphere of radius, 2\arcmin, and the outer one by a sphere of radius, 10\arcmin. 
The velocity is assumed to decrease linearly from 12\,\kms ~at the inner limit, to 2\,\kms 
~at the outer limit. 

The emission associated to Comp. 2 is clearly red-shifted with respect 
to the stellar radial velocity. This can be understood with material flowing within 
a hemisphere (Source (2)) whose axis is orientated in the direction opposite to 
the observer (i\,=\,$-$90$^{\circ}$; see Fig.~\ref{ModdemispherVconst}). The inner 
radius is set at 0.1\arcmin ~and the outer one at 1\arcmin; the expansion velocity 
is taken to decrease from 5~\kms ~at 0.1\arcmin, to 4~\kms ~at 1\arcmin.

For both sources the flux in atomic hydrogen is assumed to be constant. 
A fair adjustment to the data (see Fig.~\ref{fluxCentre_C_model_1_2}, 
\ref{flux5_model_1_2} and \ref{fluxC1E_C1W_C1EW_model_1_2}) 
is obtained by selecting fluxes corresponding to 0.74~10$^{-7}$\,\Msold 
~(Source (1)) and to 0.70~10$^{-7}$\,\Msold ~(Source (2)). 
With the parameters adopted for Sources (1) and (2), the times to build 
these sources are 57\,10$^3$ and 8\,10$^3$ years, resp., which translates to hydrogen 
masses of 4.2\,10$^{-3}$ and 5.6\,10$^{-4}$ \Msol. This is a factor 1.3 lower than 
the direct estimate from the map (Sect.~3), probably because our model tends to underestimate 
the flux density towards the East (Fig.~\ref{flux5_model_1_2}).

\begin{figure*}
\centering
\epsfig{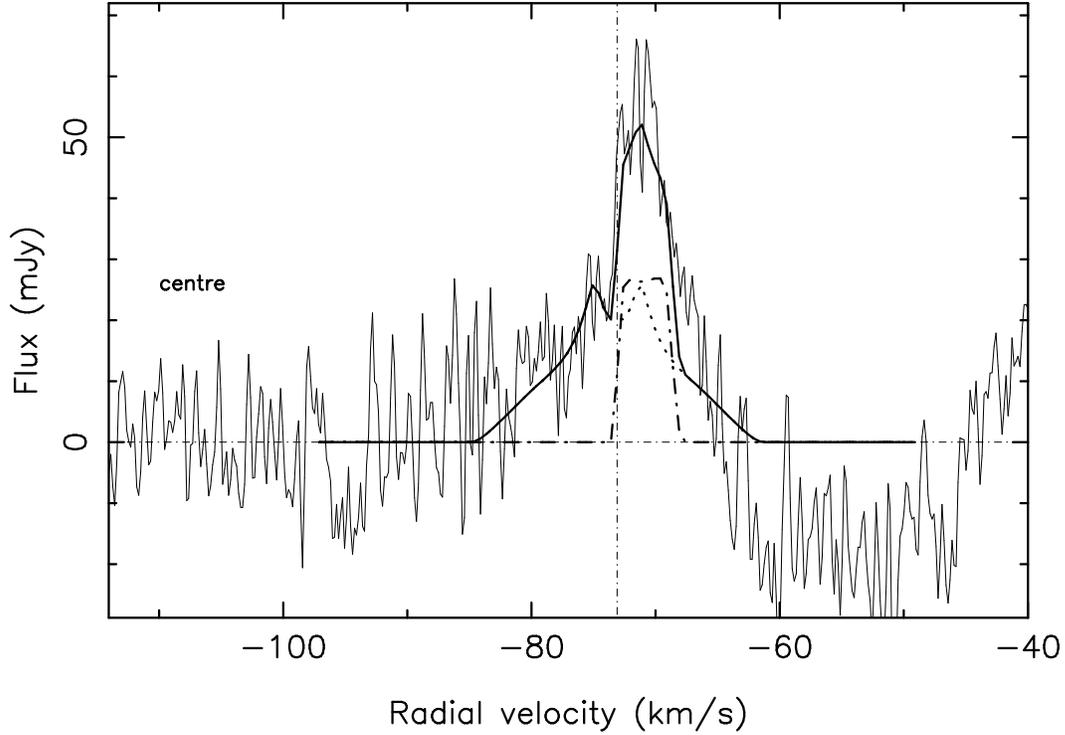}
\caption{\HI spectrum obtained with the NRT on the star position (thin line) and 
modelled spectrum (Sect. 5.2; thick line). The modelled spectrum is the sum of 
2 components\,: (i) a broad one (dotted line) produced by material flowing preferentially 
in the plane of the sky towards the North-East, and (ii) a narrow one (dash-dotted line)
produced by material flowing away from the observer. The vertical line indicates 
the star radial velocity (\Vlsr = $-$73.1 \kms).}
\label{fluxCentre_C_model_1_2}
\end{figure*}

\begin{figure*}
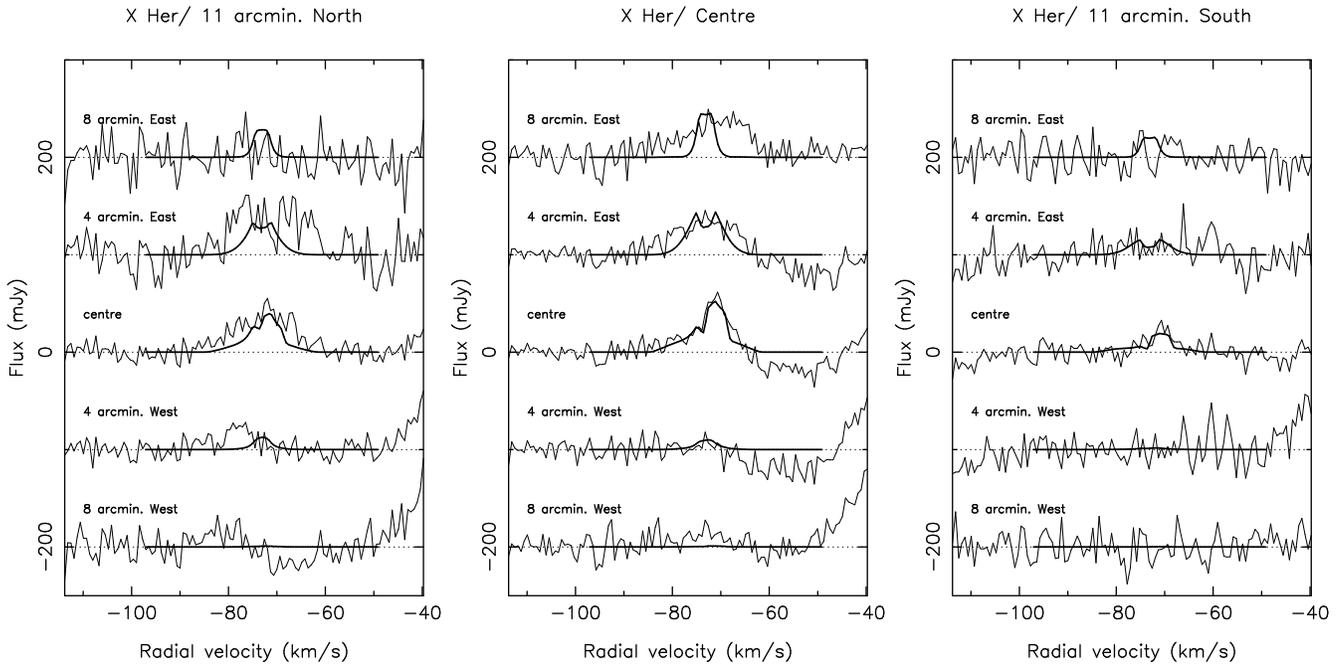

\centering
\epsfig{figure=flux5_0p5N_model_1_2_New.ps,angle=0,width=5.8cm}
\epsfig{figure=flux5_Centre_model_1_2.ps,angle=0,width=5.8cm}
\epsfig{figure=flux5_0p5S_model_1_2.ps,angle=0,width=5.8cm}
\caption{Comparison between the X Her spectra obtained at various positions 
(Fig.~\ref{Map}; thin lines) and the modelled spectra (Sect.~5.2; thick lines). 
{\it Left panel:} spectra at 11\arcmin ~North; {\it centre panel:} spectra 
at the X Her declination; {\it right panel:} spectra at 11\arcmin ~South.}
\label{flux5_model_1_2}
\end{figure*}

\section{Discussion}

The position-switch spectra presented in Sect.~3 show a compact \HI 
emission approximately centered on the X Her position and covering 
the range $-$85 to $-$60 \kms. The line-profile can be decomposed 
in 2 components: (i) a broad one (FWHM $\sim$ 13 \kms) 
centered at $-$72.2 \kms ~(Comp. 1), and (ii) a narrow one (FWHM $\sim$ 4 \kms) 
centered at $\sim$ $-$70.6 \kms ~(Comp. 2). The emission associated to Comp. 1 has a size 
of about 10$'$ and is offset to the East and to the North by about 4$'$. Comp. 2 
is not resolved spatially ($\phi \leq 4'$) and is detected only at the star position. 
We cannot strictly exclude that this emission traces a sub-structure within the halo 
of Cloud I. However, the radial velocity of Comp. 1 and its spectral extent 
closely match those of X Her in CO and SiO (Table 1), 
and Comp. 2 is, within one beam ($\pm 2'$), coincident with this star. 
Furthermore CO has been detected in the direction of X Her at a velocity 
that fits the optical one and the interferometric map obtained with BIMA 
(Nakashima 2005) shows a spatial coincidence to within a fraction of an 
arcsec. Finally an IRAS emission extended at 60 $\mu$m ( $\phi = 12'$) 
is associated to X Her (Young et al. 1993a). 
Conversely CO emission is rarely associated to HVCs and,  
with our position-switch observing procedure, 
we have now found \HI emission in the directions of more than 
20 AGB sources in the same velocity range as in CO.  Therefore, it seems 
very probable that the \HI emission detected with our position-switching 
technique is tracing 
matter belonging to the circumstellar environment of X Her 
and we adopt this viewpoint 
in the following. On the other hand, we do not consider 
that Clouds I and II are 
related to X Her (for instance through past ejection-events) 
because its proper motion is 
towards the North-West, i.e. towards them. These 2 clouds may be each 
at any distance from the Sun along closeby lines of sight and likely 
belong to the CHVC class. 

The \HI emission profile of X Her is quite comparable to those of RS Cnc (Paper I)
and EP Aqr (Paper II): a narrow component is superimposed on a broader one. The narrow 
feature is not spatially resolved by the NRT ($\phi <$4\arcmin) whereas the broad one 
is resolved. These 3 sources are M-type SRb with stellar effective temperature higher 
than 3\,000\,K, and at about the same distance ($\sim$ 130 pc). They also 
belong to the small class of sources with composite CO profiles (K1998). At this 
stage we stress that the \HI source associated to Y CVn probably belongs to a 
different class: although it shows also a narrow component superimposed on 
a broader one, the narrow component is spatially resolved while the broad one is not. 

The presence of atomic hydrogen close to these 3 central stars ($\leq$ 2~10$^{17}$\,cm) 
lends support to the GH1983 model which predicts that hydrogen should be mostly 
atomic in the atmospheres and inner envelopes of stars with effective temperature 
larger than 2\,500\,K.

It is also worth noting that these 3 sources show a silicate emission around 10 $\mu$m 
plus an unidentified dust feature around 13~$\mu$m (Speck et al. 2000; Sloan et al. 2003).
The occurrence of the 13 $\mu$m dust feature appears correlated with the existence of 
a warm CO$_2$ layer close to the central star (Justtanont et al. 1998). Our data also 
suggest a possible relation of the 13\,$\mu$m-carrier formation with a H\,{\sc {i}}-rich 
atmosphere.

A surprising characteristic is that the centroid velocities of the 2 spectral components 
do not exactly coincide. Comp. 1 is at $\sim$ $-$72.2 \kms, close to the other radio line emissions 
and to the star radial velocity (Hinkle et al. 2002). Comp. 2 is at $\sim$ $-$70.6 \kms ~and 
clearly red-shifted by $\sim$ 2-3\,\kms. In Paper II, we noted that for EP Aqr the broad (1) 
and narrow (3) components are at the same velocity, but are also shifted by $\sim$ 2-3\,\kms 
~with respect to the CO ones (K1998). For RS Cnc, one also observes a shift by 
$\sim$ $-$2\,\kms ~(Paper I and K1998). These shifts are small but real and should be 
explained, especially in view of the widths of the narrow components which are of the 
same order. For instance, in the case of EP Aqr, the narrow \HI and CO components basically 
do not overlap. The case of X Her is particularly interesting because, thanks to the monitoring 
of CO lines in the near-infrared range by Hinkle et al. (2002), the star radial velocity is known 
accurately, and because the 2 \HI velocities differ. In these conditions, the velocity shift 
of Comp. 2 can only be explained by matter flowing preferentially away from the observer with 
respect to the central star. 

The asymmetry in the \HI brightness distribution confirms that matter from X Her is flowing  
in preferred directions. It suggests that aspherical outflows may develop on large scale early 
in the AGB phase, well before the planetary nebula phase, at variance to the common vision 
(e.g. Sahai et al. 2003). X Her is probably a young AGB given the absence of technetium 
mentioned earlier.

The modelling of the \HI emission that we have developed supports these interpretations 
of the 2 spectral components observed in X Her. Assuming 75 \% of the mass in atomic 
hydrogen, we find a circumstellar mass of $\sim$ 6.4 10$^{-3}$ \Msol, of which Source (1) 
accounts for the main part. This estimate is in good agreement with that 
obtained by Young et al. (1993b) from IRAS data at 60 $\mu$m (they find 
0.009 \Msol ~at a distance of 220 pc which translates to $\sim$\,0.004\,\Msol ~at 140 pc). 
The velocity laws that we have adopted are somewhat arbitrary. 
The fitting of the quasi-Gaussian profile of Comp. 1 requires a velocity decreasing outwards 
(Paper II). This may result from a succession of mass loss episodes with velocity increasing 
with time, although it is hard to avoid an overshooting of the outer shells by the inner ones.
A more likely explanation is the interaction of the stellar wind with the ISM 
(Young et al. 1993b). For Comp. 2, as the corresponding source is not resolved, 
we cannot distinguish between a negative velocity gradient and a positive one.
Finally, we find that the two different episodes of mass loss develop 
at about the same rate, and are separated by a lapse of about 3\,000 years, 
but the duration of this lapse is only weakly constrained. 
A better spatial resolution would certainly help to constrain these velocity laws 
as well as constrain the variation of the mass loss rate as a function of time. 

The mass loss rates that we find for the 2 episodes correspond to $\sim$ 10$^{-7}$ \Msold. 
These estimates 
compare well with those obtained from the modelling of the broad CO components (K1998) 
and from IRAS data at 60 $\mu$m (Young et al. 1993b). It may be surprising that, in the end, 
the \HI narrow component does not appear to be related to the CO narrow components. 
In fact one should note that they only partially overlap in velocity. 
Also the NRT \HI beam (4\arcmin$\times$22\arcmin) is much larger than the beams 
used for CO (e.g. 30\arcsec ~for CO(2-1) and 20\arcsec ~for CO(3-2), K1998). 
Furthermore the CO photo-dissociation diameter is only $\sim$10\arcsec ~(K1998). 
KJ1996 note that ``drastic variations of line shapes from one position to another indicate 
that this envelope presents a complex small scale velocity structure''.
Therefore the materials 
responsible for these narrow components are certainly distinct. Nevertheless they could still 
belong to the same kinematical structure. However our modelling tends to indicate that it may not 
even be the case. Finally the recent work of Nakashima (2005) who suggests that the narrow 
CO(1-0) component traces in fact a disk in Keplerian rotation rather than an outflow brings 
further indication that the narrow CO and \HI components are not related. 

The narrow \HI component with a preferred direction at i\,$\approx$\,$-$90$^{\circ}$ could be 
associated to the bipolar flow detected in CO(2-1) by KJ1996. They find that it should be inclined 
at a small viewing angle corresponding to i\,$\approx$\,$-$75$^{\circ}$. 
The projection on the sky of 
the red-shifted cone should be at a position angle $\approx$\,60$^{\circ}$ (Nakashima 2005).
It is therefore of interest to consider the possibility that Source (2) has the same 
orientation as the bipolar CO flow. We performed a modelling where Source (2) is a 
hemisphere as in Sect.~5.2 (r$_{\rm in}$=0.1\arcmin, r$_{\rm out}$=1\arcmin), 
but with i\,=\,$-$75$^{\circ}$ and PA\,=\,60$^{\circ}$ (Fig.~\ref{fluxCentre_C_model_1_2_CO}). 
The fit that we obtain 
is still satisfactory, but we need to keep a maximum velocity at $\approx$ 5 \kms, 
which is much less than the velocity estimated by KJ1996 for the CO bipolar flow 
($\approx$\,10\,\kms). Also the fit to the position-switch spectra is somewhat degraded. 
We conclude that Source (2) may have approximately the same orientation 
axis as the bipolar flow detected in CO. 

\begin{figure}
\centering
\epsfig{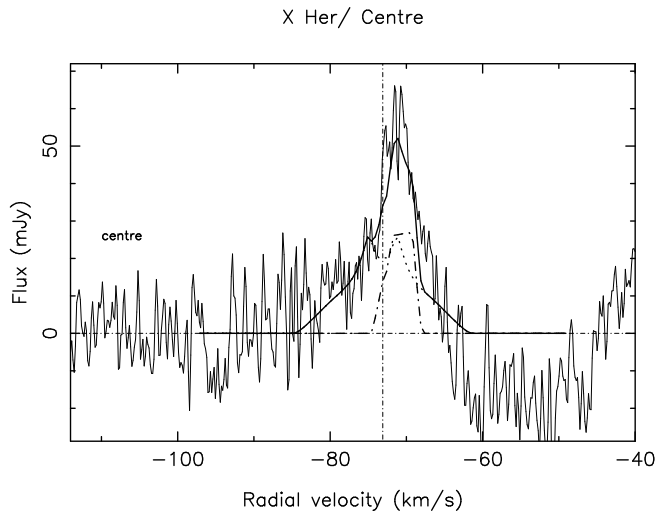}
\caption{Same as in Fig.~\ref{fluxCentre_C_model_spher}, but the second component 
is produced by material flowing within a hemisphere orientated at i\,=\,$-$75$^{\circ}$ 
and PA\,=\,60$^{\circ}$ (Sect.~6).}
\label{fluxCentre_C_model_1_2_CO}
\end{figure}

Finally it is interesting to consider the possibility that the orientation axis of Source (1)
would be closer to the CO bipolar one than in our model. Although they favor a model with 
a small viewing angle, KJ1996 admit that this viewing angle could be increased up to 45$^{\circ}$. 
To test the sensitivity of Comp. 1 to the inclination with respect to the plane of the sky 
we have increased i from 0$^{\circ}$ to $-$45$^{\circ}$. However, acceptable fits can be obtained 
only up to $\approx -$15$^{\circ}$. We conclude that the preferred direction of Source (1) is 
different from that of Source (2). 

This could be an effect of the mass being ejected 
from the star in different directions at different epochs.
Another interesting possibility, which is suggested by the large radial velocity of X Her, 
is that the stellar outflow is distorted by the ram pressure from the surrounding ISM 
in a direction close to the plane of the sky, at PA $\sim$ 45$^{\circ}$. 
We note that in our model, the \HI density at the outer 
border (10\arcmin) is $\sim$ 1.9\,cm$^{-3}$, within the range expected for the ambient  
ISM ($\approx$ 0.1--10~cm$^{-3}$).

Although the CO results have been a guide for a part of our modelling it appears that the CO 
and \HI data are complementary 
and that the \HI maps can provide valuable informations on the outer shells of red giants. 

\section{Conclusions}
   
The \HI line at 21 cm has been detected in the position-switch mode 
with the NRT in the direction of the mass-losing 
late-type giant X Her within the expected velocity range. 
Although a coincidence with an ISM cloudlet cannot be excluded, the emission is  
most likely associated with the circumstellar environment of X Her. 
It is spatially resolved ($\phi \sim 10'$). 
As there is moderate interstellar confusion this source is appropriate 
for a detailed study of circumstellar \HI emission.

The profile is composite with a narrow component superimposed on a broader one. 
The narrow component is not resolved spatially ($\phi <$ 4\arcmin) while 
the broad one is extended. These properties are similar to those observed on EP Aqr 
and RS Cnc, two sources which share with X Her many other common properties. 

The two spectral components are centered at slightly different velocities, which is 
a strong indication that the mass loss is not spherically symmetric. The spatial distribution 
of the \HI brightness also points to a non-symmetric geometry.

Our spatially resolved \HI data can be modelled with 2 sources: (i) a flow in a direction 
close to the plane of the sky whose properties match approximately 
those obtained from the IRAS data at 60 $\mu$m, 
(ii) a second flow within a hemisphere opposite to the observer that may be related 
to the bipolar flow observed in CO. 
The masses of atomic hydrogen associated with these two components 
are $\sim 4~10^{-3}$ and 6~10$^{-4}$ \Msol, respectively. 
The HI data probe the circumstellar shell of X Her over a large region ($\sim$~0.4 pc) 
that has been filled during a long time ($\sim$ 10$^5$ years). During this long period the 
geometry of the outflow has probably changed significantly. 
Finally, the interaction of the stellar wind with the ambient 
ISM may affect the \HI spatial distribution as well as the spectral profiles.

The total \HI mass and production rates measured here are in agreement with those 
deduced more indirectly from CO and IRAS data. The \HI and IRAS 60 $\mu$m angular 
extents are comparable, although the IRAS source size could have been limited by the 
dust temperature gradient. 

More generally, our data illustrate the need of a large spectral resolution 
($\sim$ 10$^6$), that is provided by the heterodyne technique, 
for describing the geometry and the kinematics of late-type giant outflows. 
Furthermore a better imaging, with a finer spatial resolution, would also 
be essential to reconstruct the history of mass loss over the past 10$^5$ years. 

\section*{Acknowledgments}

We acknowledge stimulating discussions with Dr. J.M. Winters. 
We thank our referee, Prof. W.B. Burton, for insightful comments. 
The Nan\c{c}ay Radio Observatory is the Unit\'e scientifique de Nan\c{c}ay of 
the Observatoire de Paris, associated as Unit\'e de Service et de Recherche 
(USR) No. B704 to the French Centre National de la Recherche Scientifique 
(CNRS). The Nan\c{c}ay Observatory also gratefully acknowledges the financial 
support of the Conseil R\'egional de la R\'egion Centre in France. This 
research has made use of the SIMBAD database, operated at CDS, Strasbourg, 
France and of the NASA's Astrophysics Data System.

\label{lastpage}

\end{document}